 \documentstyle[twocolumn,prl,floats,aps,epsf]{revtex} % epsfig
 \newcommand{\figwidth}{3.375in} % Use for two-column output

\begin{document}
\preprint{UBCTP-95-003}
\draft

% comment out the following line for single column output
%
\twocolumn[\hsize\textwidth\columnwidth\hsize\csname @twocolumnfalse\endcsname

\title{Phase Transitions in the One-Dimensional Pair-Hopping Model:\\
a Renormalization Group Study}
\author{Arnold~E.~Sikkema$^a$ and Ian~Affleck$^{a,b}$}
\address{$^a$Physics Department, University
of British Columbia, Vancouver, BC, V6T~1Z1, Canada}
\address{$^b$Canadian Institute for Advanced Research, University of
British Columbia, Vancouver, BC, V6T~1Z1, Canada }
\date{Received by {\sl Phys. Rev.} {\bf B} on 20 April 1995}
\maketitle

\begin{abstract}
\noindent
The phase diagram of a one-dimensional tight-binding model with a
pair-hopping term (amplitude $V$) has been the subject of some
 controvery. Using
two-loop renormalization group equations and
the density matrix renormalization group with
lengths $L\leq 60$, we argue that no spin-gap transition occurs at
half-filling for positive $V$, contrary to recent claims.  However, we
point out that away from half-filling, a {\it phase-separation}
 transition occurs at finite $V$.  This transition and the spin-gap
transition occuring at half-filling and {\it negative} $V$ are analyzed
numerically.
\end{abstract}
\pacs{PACS Numbers: 75.20.Hr\hfill Preprint \# UBCTP-95-003, cond-mat/9505006}

% comment out the following line for single column output
%
]

\section{Introduction}

In recent years interest in correlated electron systems has increased
particularly in an attempt to understand high-temperature
superconductors.  It is important to study the effect of all possible
nearest-neighbour interactions; however, the hopping of on-site
spin-singlet pairs has not been well studied to date.

The Hamiltonian for the pair-hopping model is\cite{Penson}
\begin{equation}
H = - t \sum_{\langle ij\rangle\sigma}[ c_{i\sigma}^\dagger c_{j\sigma} +
\hbox{h.c.}]
    - V \sum_{\langle ij\rangle} [c_{i\uparrow}^\dagger c_{i\downarrow}^\dagger
c_{j\downarrow} c_{j\uparrow}+\hbox{h.c.}] \label{Ham}
\end{equation}
where $c_{i\sigma}^\dagger$  ($c_{i\sigma}$) creates (destroys) an
electron of spin $\sigma=\uparrow,\downarrow$ at lattice site $i$, so
that $n_{i\sigma}=c_{i\sigma}^\dagger c_{i\sigma}$ is the number of
electrons of spin $\sigma$ at site $i$, and $\langle ij\rangle$ denotes
nearest-neighbour pairs.  Thus $t$ and $V$ are the single-electron- and
pair-hopping amplitudes respectively, so this models a competition
between the two hopping terms.  As $t\to -t$ is a symmetry of $H$, in
this paper we take $t\ge 0$.

At large $|V|/t$, all sites are doubly occupied or empty (assuming an
even number of electrons) and the model becomes equivalent to spinless
fermions\cite{Penson}.  In particular there is a large gap, of
${\cal O}(V)$, to any excited state with non-zero spin.  This is true
for either sign of $V$, but it is important to note that $V\to -V$ is
{\it not} a symmetry of the model, unlike the Hubbard model.
Finite-size numerical work\cite{Penson,Hui} has been performed on this
model in one dimension for positive $V$, suggesting a phase transition
at which the spin gap (or single-particle excitation gap) opens, at
$V/t\approx 1.4$.  Two different analytical renormalization group (RG)
analyses have been applied to the model. One\cite{Affleck} suggested
the existence of a spin gap $\Delta_{\rm s}$ for all $V>0$, with
\begin{equation}
\Delta_{\rm s} \propto {\rm e}^{-\pi t/V}
\end{equation}
as $V/t\to 0$, and no transition for any positive $V$; the
other\cite{Hui} suggested that there is a transition at
$V/t\approx 1.4$, consistent with the numerical work.

One purpose of this paper is to re-examine this question.  Previous
numerical work has used chains of length $L\leq 12$.  We present data
for much longer chains, $L\leq 60$, using White's density matrix
renormalization group (DMRG) technique\cite{White}, thereby countering
the dominance of finite-size effects.  We also discuss the subtleties
involved in trying to extract information about the phase diagram from
a low-order analytical  RG calculation.  Our conclusion is the same as
that of Ref.~\cite{Affleck}: no phase transition for any $V>0$.

It was argued in Ref.~\cite{Affleck} that there {\it is} a phase
transition, corresponding to the appearance of a spin gap, for some
finite {\it negative} $V$.  (The case $V<0$ was not studied in
Refs.~\cite{Penson} or \cite{Hui}.) We analyze this numerically,
finding a transition at $V_{\rm c}\approx -1.5t$.

There has been considerable interest of late in phase separation in the
Hubbard and $t$-$J$ models in one and higher dimensions.  The
pair-hopping model provides a simple example of a model where it is
easy to see that a phase-separation transition must occur at some
finite critical coupling, with a non-zero total magnetization.
Consider the model at $t=0$, with a total magnetization $M/2$,
corresponding to an excess of $M$ spin-up electrons.  These electrons
necessarily reside on singly occupied sites and are therefore
completely immobile when $t=0$.  Thus the model is equivalent to
spinless fermions with vanishing hopping terms to $M$ sites.
Equivalently we have an $XY$ spin chain with vanishing exchange
coupling to $M$ sites, corresponding to $M$ nonmagnetic impurities.
These simply have the effect of breaking the chain up into chainlets.
It is fairly clear, and can be demonstrated explicitly by a trivial
calculation in the free spinless fermion model, that the energy is
lowest when all $M$ impurities are next to each other, leaving an $XY$
chain of $L-M$ sites.  This corresponds to phase separation at
$|V|/t\gg 1$ for any non-zero magnetization. On the other hand, the
renormalization group analysis of the model at weak coupling,
$|V|/t\ll 1$, indicates behaviour similar to that of the Hubbard model,
with no phase separation.  This suggests that  phase-separation
transitions should occur at finite values of $V/t$ (one for positive
$V$ and one for negative $V$).  We find  evidence for such a transition
at $V=V_{\rm c1}\approx 3.5t$, but we have not examined the one at
$V=V_{\rm c2}<0$.

In Sec. \ref{sec:RG} we review and critically analyze the previous
analytical RG calculations.  In Sec. \ref{sec:largeV} we analyze
quantitatively the large $|V|/t$ limit.  In Sec. \ref{sec:posV} we
present our numerical work at half-filling, $M=0$ and $V>0$, indicating
no phase transition.  In Sec. \ref{sec:negV} we present numerical
evidence for the spin-gap transition at half-filling, $M=0$ and $V<0$.
In Sec. \ref{sec:phase-sep} we present numerical evidence for the
phase-separation transition for $M\neq 0$ at $V=V_{\rm c1}\approx 3.5t$.

\section{Analytical RG Studies of the Phase Diagram}
\label{sec:RG}

The RG analyses of Refs.~\cite{Affleck} and \cite{Hui} came to quite
different conclusions.  Here we would like to explain the reasons for
this and give arguments in favour of the former approach.

We use essentially the notation of Ref.~\cite{Hui}, which is taken from
the review article of S\'olyom\cite{Solyom}.  Taking the continuum
limit of the pair-hopping model we obtain a general Hamiltonian:
\begin{equation}
H = \int {\rm d}x[{\cal H}_0+{\cal H}_{\rm int}],
\end{equation}
where ${\cal H}_0$ and ${\cal H}_{\rm int}$ are the dimensionless
kinetic energy density and interaction Hamiltonian density.  We keep
only wave vectors close to the two Fermi points,
$k\approx \pm k_{\rm F}$, which we label, in position space,
$\psi_{\rm L}$ and $\psi_{\rm R}$.  The kinetic energy density is given
by
\begin{equation}
{\cal H}_0 = v_{\rm F}
[\psi^{\alpha\dagger}_{\rm L}
{\rm i}{{\rm d}\over {\rm d}x}\psi_{{\rm L}\alpha}
 -\psi^{\alpha\dagger}_{\rm R}
{\rm i}{{\rm d}\over {\rm d}x}\psi_{{\rm R}\alpha}].
\end{equation}
Here the spin index $\alpha$ is implicitly summed over; $v_{\rm F}$ is
the Fermi velocity.  Some of the various interaction terms can be
conveniently written in terms of charge and spin currents (or
densities)
\begin{equation}
J^\rho_{\rm L}\equiv \psi^{\alpha\dagger}_{\rm L}\psi_{{\rm L}\alpha}
{\rm~and~}
\vec J^{\rm s}_{\rm L}\equiv
\psi^{\alpha\dagger}_{\rm L}
{\vec \sigma^\beta_\alpha \over 2}
\psi_{{\rm L}\beta},
\end{equation}
and similarly for $J_{\rm R}^\rho$ and $\vec J_{\rm R}^s$.  In this
notation, we have
\begin{eqnarray}
{\cal H}_{\rm int}&=&\pi v_{\rm F}\left\{-{1\over 2}g_\rho
J^\rho_{\rm L}J^\rho_{\rm R}
-2g_{\rm s}\vec J^{\rm s}_{\rm L}\cdot \vec J^{\rm s}_{\rm R}\right.\nonumber\\
&&\:\:\:\:\:\:\:\:\:
-{g_3\over 4}
[\epsilon_{\alpha \beta}\psi^{\alpha \dagger}_{\rm L}
\psi^{\beta \dagger}_{\rm L}
\epsilon^{\gamma \delta}\psi_{{\rm R}\gamma}\psi_{{\rm R} \delta}
+({\rm L}\leftrightarrow {\rm R})]\nonumber \\
 &&\:\:\:\:\:\:\:\:\:\left.
-{g_4\over 4}[J^\rho_{\rm L}J^\rho_{\rm L}
-{4\over 3}\vec J^{\rm s}_{\rm L}\cdot \vec J^{\rm s}_{\rm L}
+({\rm L}\leftrightarrow {\rm R})]\right\}.
\label{HSol}
\end{eqnarray}
We have chosen to write the Hamiltonian in a manifestly $SU(2)$
invariant way.  The last term can also be written as
\begin{equation}
J^\rho_{\rm L}J^\rho_{\rm L}
-{4\over 3}\vec J^{\rm s}_{\rm L}\cdot \vec J^{\rm s}_{\rm L}
=J^\rho_{\rm L}J^\rho_{\rm L}
-4J^{{\rm s}z}_{\rm L}J^{{\rm s}z}_{\rm L}
=4J_{{\rm L}\uparrow}J_{{\rm L}\downarrow},
\end{equation}
where
\begin{equation}
J_{{\rm L}\alpha }=\psi^{\alpha \dagger}_{\rm L}\psi_{{\rm L}\alpha}
{\rm~(repeated~index~not~summed)}.
\end{equation}
To the first non-vanishing order in $V$, the bare couplings have the values
\begin{equation}
v_{\rm F}=2t;\:
g_\rho=
-g_{\rm s}=
g_3=
g_4=2V/\pi v_{\rm F}.
\label{bare}
\end{equation}
To cubic order, the RG equations are given by
\begin{eqnarray}
-{{\rm d}g_{\rm s}   \over {\rm d}l}
&=&g_{\rm s}^2+\textstyle{1\over 2}(g_{\rm s}+g_4)g_{\rm s}^2\nonumber \\
-{{\rm d}g_\rho\over {\rm d}l}
&=&g_3^2+\textstyle{1\over 2}(g_\rho -g_4)g_3^2\nonumber \\
-{{\rm d}g_3   \over {\rm d}l}
&=&g_\rho g_3 +\textstyle{1\over 4}(g_\rho^2+g_3^2-2g_\rho g_4)g_3\nonumber \\
-{{\rm d}g_4\over {\rm d}l}
&=&\textstyle{3\over 4}(g_\rho g_3^2-g_{\rm s}^3).
\label{RG}
\end{eqnarray}
Here $l=-\log \Lambda$, where $\Lambda$ is an ultraviolet cut-off.  As
we lower the cut-off to study the long-distance behaviour, $l$
increases.

Part of the discrepancy between the conclusions of Ref.~\cite{Affleck}
and Ref.~\cite{Hui} arises from the treatment of the $g_4$ coupling.
If we bosonize the theory, then
\begin{eqnarray}
J^\rho_{\rm L}J^\rho_{\rm L}
&=&{1\over 2\pi}(\partial_+\phi_\rho)^2 \nonumber \\
4J^{{\rm s}z}_{\rm L}J^{{\rm s}z}_{\rm L}
&=&{1\over 2\pi}(\partial_+\phi_{\rm s})^2,
\end{eqnarray}
where $\phi_{\rho,{\rm s}}$ are charge and spin bosons.  Hence $g_4$ simply
shifts the velocities of charge and spin excitations to
\begin{eqnarray}
v_\rho&=&v_{\rm F}(1+g_4/2)\nonumber \\
v_{\rm s}&=&v_{\rm F}(1-g_4/2).
\end{eqnarray}

A common approach to Luttinger liquids is to simply set $v_\rho$ and
$v_{\rm s}$ to their renormalized values and drop $g_4$ from the RG
equations.  This approach was used in Ref.~\cite{Affleck}.  The RG
equation for $g_{\rm s}$ then decouples from the $g_\rho$ and $g_3$
ones.  This arises from the fact that, upon bosonizing, the
corresponding operators involve only the spin boson and only the charge
boson respectively.  We then see that $g_{\rm s}=0$ is not a stable
fixed point:  if $g_{\rm s}<0$, as is the case for $V>0$, $g_{\rm s}$
will flow away to strong coupling.  This is usually taken to indicate
that the system is in a phase with a gap for spin excitations.

On the other hand, a quite different conclusion can be reached if $g_4$
is kept in the RG equations.  Then, according to Eq.~(\ref{RG}),
$g_{\rm s}=0$ becomes a stable fixed point from the negative side
provided that $g_4<-2$.  The nature of this putative phase can be
understood by also rewriting the free electron kinetic energy in terms
of spin and charge currents.  Setting all coupling constants to zero
except $g_4$, the full Hamiltonian density can be written
\begin{equation}
{\cal H} =
{\pi v_{\rm F}\over 2}
\left[
 \left(1-{g_4\over 2}\right) J^\rho_{\rm L}J^\rho_{\rm L}
+\left(1+{g_4\over 2}\right) J^{{\rm s}z}_{\rm L}J^{{\rm s}z}_{\rm L}
\right] +
({\rm L}\leftrightarrow {\rm R}).
\label{Sugawara}
\end{equation}
We see that for $g_4<-2$, the spin part of the Hamiltonian becomes
unstable. That is, $J^{{\rm s}z}_{\rm L}(x)$ and $J^{{\rm s}z}_{\rm R}(x)$
tend to become large, necessitating the keeping of
higher order terms in the
Hamiltonian. On the other hand, the condition of zero total
magnetization requires
\begin{equation}
\int {\rm d}x[J^{{\rm s}z}_{\rm L}+J^{{\rm s}z}_{\rm R}]=0.
\end{equation}
A possible interpretation of this phase (which occurs in other known
cases) is a ferromagnetic phase.  The condition of zero total
magnetization forces a domain structure, i.e.\ phase separation, to
occur.  One side of the system has positive polarization and the other
half negative.

In Ref.~\cite{Hui}, the cubic RG equations were integrated, including
$g_4$, using the initial values of Eq.~(\ref{bare}) [plus the
${\cal O}(V^2)$ corrections which are not important at small V].  The
result was that for $0<V/t<1$, a fixed point was reached with
$g_4\approx -2.5$ and $g_{\rm s}=0$.

Whether or not $g_4$ is included, for $V/t<1$, $g_3$ renormalizes to
zero and $g_\rho$ to some small positive value which
depends on $V/t$,
corresponding to zero gap for charge excitations.

In Ref.~\cite{Affleck} this phase was identified as having a spin gap
since $g_{\rm s}$ does not flow to zero.  In Ref.~\cite{Hui} this
phase, with $g_{\rm s}=0$ and $g_4<-2$, was assumed to have no gap for
single-particle excitations.  Since these excitations have spin
${1\over 2}$ and charge 1 this would imply, from the usual Luttinger
liquid viewpoint that there is neither a charge gap nor a spin gap.  We
do not find this calcuation convincing.  It is not possible to argue
rigorously that $g_4$ renormalizes to a value less than $-2$ using
only the cubic order RG equations.  If this actually happened, as
claimed in Ref.~\cite{Hui}, this would presumably imply a transition
into a ferromagnetic phase (or possibly some other more exotic phase
characterized by the harmonic spin Hamiltonian of Eq.~(\ref{Sugawara})
becoming unstable) for arbitrarily small $V$.  No direct numerical
evidence for ferromagnetism (or other exotic behaviour) at small $V/t$
has been presented.  Although earlier numerical work in
Refs.~\cite{Penson} and \cite{Hui} saw indications of vanishing spin
gap in this region of parameters, the numerical results presented here
in Sec.~\ref{sec:posV} based on much longer chains ($L\leq 60$ instead
of $L\leq 12$) find a non-zero spin gap.

\begin{figure}
\epsfxsize=\figwidth                                             % epsfig
\centerline{(a)}\centerline{\epsffile{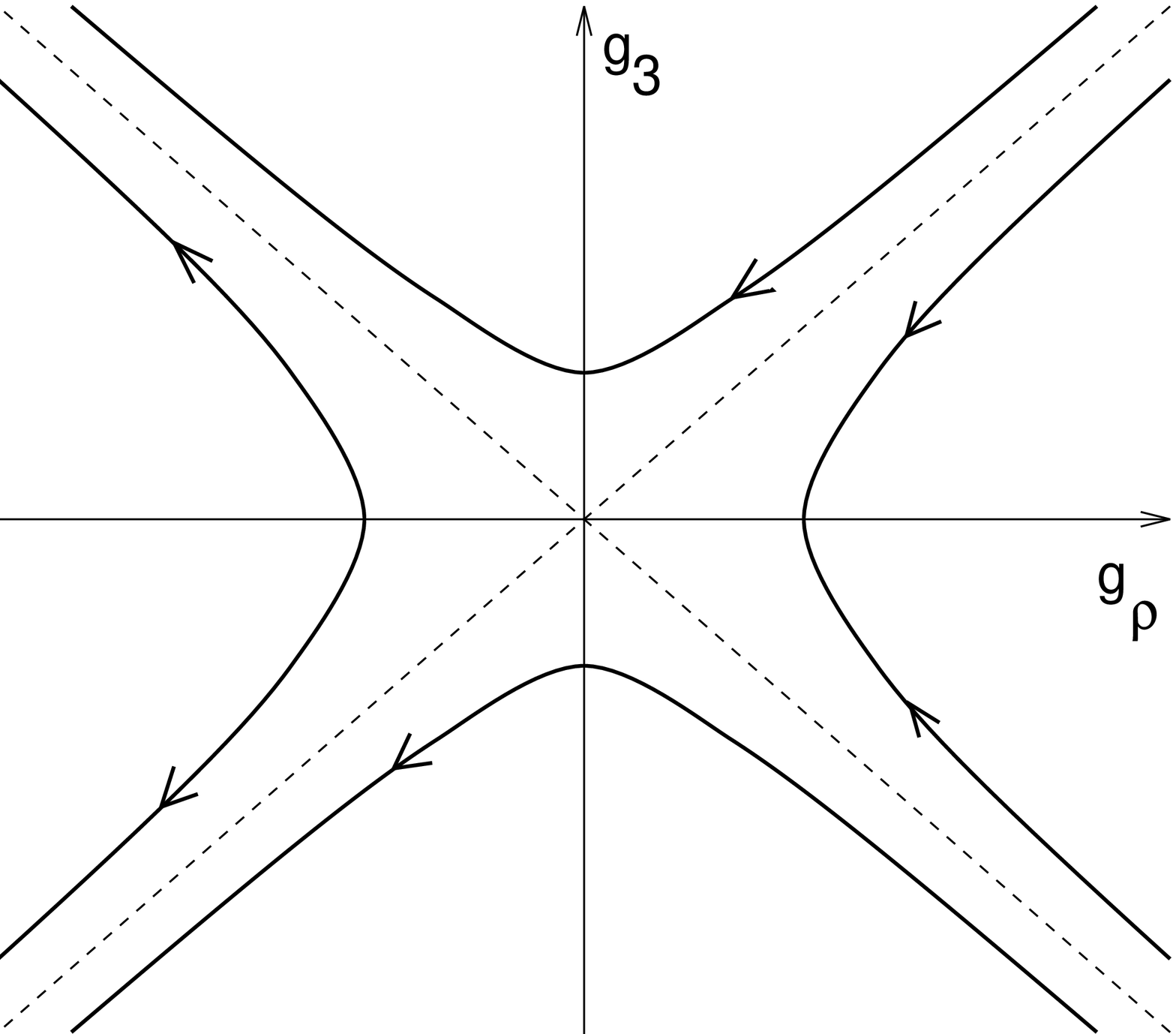}}\vspace{3mm}  % epsfig
\centerline{(b)}\vspace{2mm}\centerline{\epsffile{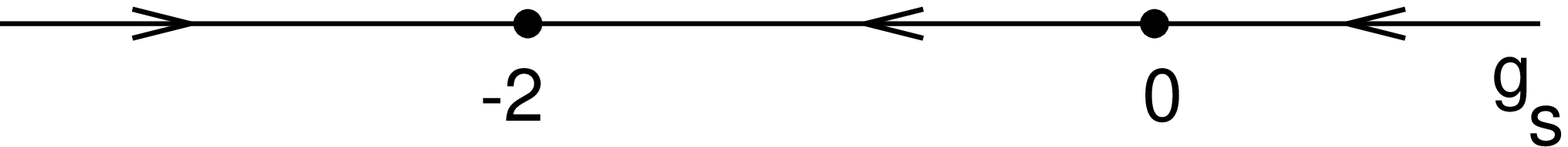}}  % epsfig
\caption{Third-order RG flow diagrams, ignoring $g_4$.\newline
(a) Flow in the charge sector, for small $g_3$ and $g_\rho$.\newline
(b) Flow in the spin sector. }
\label{fig:flows}
\end{figure}

In Ref.~\cite{Hui} a different phase is reached for $V/t>1$ with a
non-zero $g_{\rm s}$ at the fixed point, corresponding to a spin gap as in
Ref.~\cite{Affleck}.  However Refs.~\cite{Affleck} and \cite{Hui} now
disagree about the behaviour of the charge couplings, $g_\rho$ and
$g_3$.  Note that the second and third RG equations in Eq.~(\ref{RG}) imply
that $g_3=\pm g_\rho$ are separatrices (for $g_4=0$).  For $g_\rho>0$,
if $|g_3|\leq g_\rho$, $g_3$ flows to zero (see Fig.~\ref{fig:flows}),
corresponding to a harmonic gapless effective Hamiltonian for charge.
Outside this region both $g_3$ and $g_\rho$ flow off to values of
${\cal O}(1)$. This is normally interpreted as a phase with a charge
gap. It is a remarkable feature of the pair-hopping model that, to
${\cal O}(V)$, $g_\rho=g_3$: the system lies on a separatrix.  It is
necessary to calculate the bare couplings to ${\cal O}(V^2)$ to deduce
whether or not $g_3$ flows to zero.  Both papers agree that these
${\cal O}(V^2)$ terms place the bare couplings in the basin of
attraction of the $g_3=0$ critical line, for small $V$.  In
Ref.~\cite{Affleck} it was assumed (on the grounds of simplicity) that
the system remained in this basin of attraction for all $V>0$.  On the
other hand, in Ref.~\cite{Hui}, the expression for the bare couplings
to ${\cal O}(V^2)$ was used for arbitrarily large $V$ to deduce that
the bare couplings moved outside this basin of attraction at a critical
$V\approx t$ (the same critical point at which $g_{\rm s}$ and $g_4$
change).  The cubic RG equations predict a fixed point at
$g_\rho=g_3=-2$, which the authors of Ref.~\cite{Hui} assume
corresponds to vanishing charge gap.

This argument concerns us because it is not possible to tell from these
low order calculations of the bare couplings and the RG equations
whether or not the bare couplings ever leave the domain of attraction
of the $g_3=0$ critical line.  Furthermore, if they did, this phase
would normally be identified as having a charge gap, which we know does
not occur for small or large $V$.  (The existence of an apparent,
finite coupling fixed point of the cubic RG equations at couplings of
${\cal O}(1)$ does not necessarily signal the existence of a different
critical point.  It could disappear upon keeping higher order terms.)

In Sec.~\ref{sec:largeV} we give analytic arguments implying that, for large
$|V|/t$, there is a spin gap but no charge gap.

By ignoring $g_4$ (i.e.\ absorbing it into velocity renormalizations) and
making a plausible assumption about the behaviour of bare coupling
constants at large $V$, we obtain simple behaviour requiring no phase
transition for any $V>0$. There is always a spin gap and no charge gap.

On the other hand\cite{Hui}, by including the renormalization of $g_4$
and using weak coupling results at strong coupling one obtains two
different phases: a bizarre small $V$ phase with an unstable harmonic
spin Hamiltonian and a large $V$ phase which would likely correspond to
a charge gap, in contradiction with the expected large $V$ result.

The authors of Ref.~\cite{Hui} applied the same RG analysis to the
positive $U$ Hubbard model.  Their analysis gave a small $U$ phase with
$g_4>2$, corresponding to a negative harmonic Hamiltonian in the charge
sector and a large $U$ phase with a non-zero $g_{\rm s}$ which would
normally correspond to a spin gap.  As they pointed out themselves,
this is in contradiction with the expected behaviour which is a charge
gap and no spin gap for all positive $U$ (at half-filling).

\section{Large $|V|$ Limit:  Spin Gap and Phase Separation}
\label{sec:largeV}

The pair destruction operators
$a_i\equiv c_{i\downarrow} c_{i\uparrow}$ have commutation relations
$[a_i,a^\dagger_j]=(1-n_i)\delta_{ij}$, where
$n_i=\sum_{\sigma}n_{i\sigma}$.  So the $a$'s commute if $n_i=1$ since
they have no effect on singly occupied sites.  In the other two cases
$n_i=0,2$ we have $n_i=2a^\dagger_i a_i$ so that the $a_i$'s are
spinless fermion operators, obeying $\{a_i,a^\dagger_j\}=\delta_{ij}$.
As shown in Ref.~\cite{Penson}, setting $t=0$ results in a ground state
involving only empty and doubly occupied sites, so the on-site pairs
are effectively spinless fermions.

The DMRG method that we use requires free boundary conditions.  So we
analytically examine the large $|V|$ limit for an open chain, noting
that nothing essential will change in going to periodic or
infinite length chains.  The ground state energy for an open chain of
even length $L$ is easily computed to be
\begin{equation}
E_0=-2|V|\sum_{n=1}^{L/2}\cos{n\pi\over L+1}
   =|V|\left(1-\csc{\pi\over 2L+2}\right).
\end{equation}
Adding a single electron to this half-filling ground state produces an
immobile site since $t=0$, effectively breaking the chain.  The energy
will depend on the location of the break, and is easily shown to be
minimized if the break is at the end of the chain, in which case the
energy is that of ${L\over 2}$ pairs hopping on an open chain of length
$L-1$, namely
\begin{equation}
E_1=-2|V|\sum_{n=1}^{L/2}\cos{n\pi\over L}
   = |V|\left(1-\cot{\pi\over 2L}\right).
\end{equation}
So the single-particle gap for the open chain is
\begin{eqnarray}
\Delta_{\rm sp}
&=&|V|\left[\csc\frac{\pi}{2L+2}-\cot\frac{\pi}{2L}\right]\\
&=&
|V|\left[\frac{2}{\pi} + \frac{\pi}{4L} - \frac{\pi}{12L^2} +
{\cal O}\left(\frac{1}{L^3}\right)\right].
\end{eqnarray}
So for $t=0$, we have a model equivalent to free spinless fermions,
corresponding to a spin gap proportional to $|V|$ but no charge gap.
To see whether this situation persists for finite $|V|/t$, we can do
perturbation theory in the lattice model in $t/V$.  This is very
similar to the well-known results on the large-$U$ Hubbard model.  In
this case we project out singly occupied sites.  A single application
of $t$ takes us into the high-energy subspace with 2 singly occupied
sites.  In second order perturbation theory we generate an effective
interaction of ${\cal O}(t^2/V)$ in the spinless fermion model.  This
simply corresponds to a nearest-neighbour interaction of the spinless
fermions.  This interaction is known to be exactly marginal, leading to
a critical line with vanishing gap.

Thus there is a spin gap for large $|V|/t$.  As there is no spin (or
charge) gap for $V=0$, there must be some transition.  On the basis of
a reliable interpretation of the analytical RG equations and careful
consideration of and comparison with numerical RG results, we conclude
that the positive $V$ transition occurs at $V=0$ instead of at some
finite $|V|$.  We show numerically that for small positive $V$, the
behaviour of the single-particle gap is of the form predicted by the RG
flows (upon dropping $g_4$) in the numerically accessible region of
phase space.  We also find a spin-gap transition at
$V=V_{\rm c}\approx -1.5t$.

The above analysis also shows that in the case $t=0$, a single unpaired
electron sits at a chain end; it is clear that additional electrons of
the same $S^z$ will clump at the chain ends as well.  That is, at
finite magnetization, the chain phase separates:  one part of the chain
assumes the net magnetization.  It is important to note that this is
not a peculiarity of the open chain; in the periodic case as well, at
$t=0$, added polarized electrons cut the chain and the chain-breaking
energy is clearly minimized by clumping them together.  Since going
 from $t=0$ to some large but finite $|V|/t$ introduces only a marginal
operator, it is clear that this phase separation will persist to some
critical values of $V$, which probably have different absolute values
because the Hamiltonian is not symmetric under $V\to -V$.

\section{Results of Numerical RG}

\subsection{DMRG Details}

We use the ``infinite system DMRG method''\cite{White}, treating open
chains of even length up to 60, and maintaining 64 (sometimes 128)
states in each block.  The ground state has total spin 0 and is at
half-filling; we add a single electron (pair of electrons of opposite
spin) to compute the spin (charge) gap for each length.  These results
are extrapolated to infinite length taking into account truncation
error uncertainties.  The figures summarize the results of our DMRG
calculations, as explained in this section.

\subsection{Spin Gap for $V>0$}
\label{sec:posV}

\begin{figure}
\epsfxsize=\figwidth\centerline{\epsffile{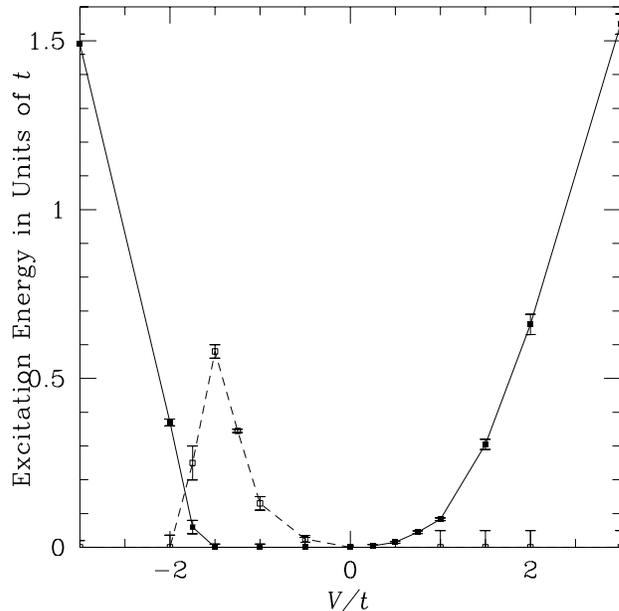}} % epsfig
\caption{Summary of numerical results:  the open squares and dashed
line are the charge gap, and the filled squares and solid line are
the spin gap.  A clear phase transition is evident near
$V=V_{\rm c}\approx -1.5t$, but for positive $V$, the spin gap opens up
 from $V=0$.  The error bars indicate uncertainty in extrapolating
$L^{-1}\to 0$; the lines are to guide the eye.}
\label{fig:gaps-all-V}
\end{figure}

We find that the spin gap does not vanish for any $V>0$, as shown in
Fig.~\ref{fig:gaps-all-V}.  In comparing its dependence on $V$ with
that predicted from the analytical RG of Ref.~\cite{Affleck}, namely
\begin{equation}
\Delta_{\rm s}\approx t{\rm e}^{-\pi t/V}
\end{equation}
(after correcting the typographical error) which is valid for small
$V/t$, the numerical work is not dependable for $V/t<1$ because there
the expected correlation length $\xi\approx v_{\rm F}/2\Delta$ becomes
of order the system size $L$.  The finite-size gap alone is
$\Delta_{\rm FS}\approx\pi v_{\rm F}/L$ so that one cannot expect to
measure $\Delta/t$ lower than
$\Delta_{\rm FS}/t\approx2\pi/L\approx0.1$ for $L=60$.

\begin{figure}
\epsfxsize=\figwidth\centerline{\epsffile{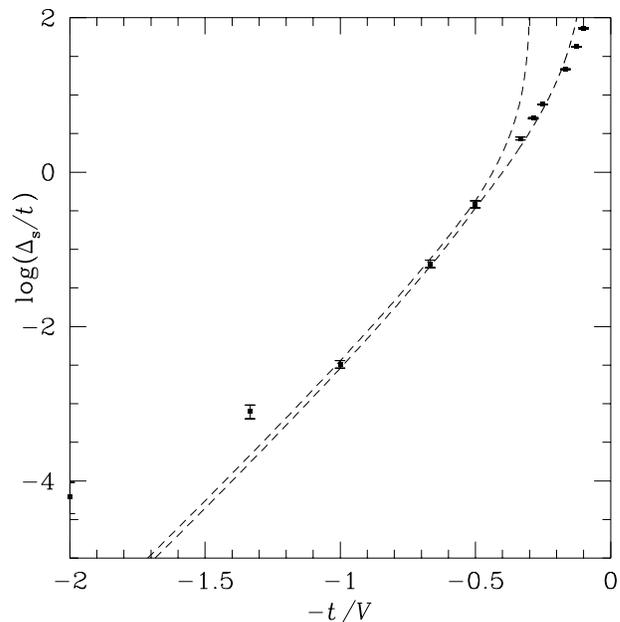}} % epsfig
\caption{Fitting the DMRG data for only two points, namely $V/t=1,2$,
to the form given by Eq.~(\protect\ref{eq:fit}), the dashed lines are
the upper and lower limits of the resulting fitted curves taking the
numerical error bars into account.  The lower limit extrapolates well
over the range $t<V<4t$, which is the expected region of validity.
(The fit is not expected to be valid for $V<t$ because of the
finite-size gap, while $V>V_{\rm c1}\approx 3.5t$ is the phase-separated
region.)}
\label{fig:fit}
\end{figure}

The RG flow equations to two-loop order, after dropping $g_4$ as
explained in Sec.~\ref{sec:RG}, give for the spin coupling
$g_{\rm s}$
\begin{equation}
g_{\rm s}^{-1}-g_{{\rm s}0}^{-1}
-{1\over 2}\log{1 +2g_{\rm s}^{-1}\over 1+2g_{{\rm s}0}^{-1}}
=\log{L\over L_0}
\end{equation}
where $L_0$ is an initial length scale ($\Lambda=L^{-1}$ is the
ultraviolet cut-off) and the initial spin coupling is\cite{Hui}
\begin{equation}
g_{{\rm s}0}=-{V\over\pi t}+\left({V\over\pi t}\right)^2
\log\left(\tan{1\over tL_0}\right).
\end{equation}
We take the spin gap to be the energy (inverse length) scale at which
$g_{\rm s}$ enters the regime of strong coupling, specifically where
$g_{\rm s}=a={\cal O}(-1)$, resulting in
\begin{equation}
\Delta_{\rm s}(V)=\Delta_0\exp(g_{{\rm s}0}^{-1}-a^{-1})
\sqrt{1+2a^{-1}\over 1+2g_{{\rm s}0}^{-1}}
\label{eq:fit}
\end{equation}
where $a={\cal O}(-1)$ is used as the criterion for $|g_{\rm s}|$
becoming large.  Fig.~\ref{fig:fit} shows that the numerically computed
spin gap is indeed of the form predicted by the RG flow equations with
$g_4$ dropped.

While the above comparison of the RG flows are to numerical results on
open (not periodic) chains, we believe these results (and in particular
the non-vanishing of the gap for $V>0$) constitute a reliable estimate
of the situation in the thermodynamic limit. However the situation is
somewhat different in the phase-separated region ($V>V_{\rm c1}\approx
3.5t$ and $V<V_{\rm c2}<0$) than in the non-phase-separated region at
smaller $|V|$.  In the non-phase-separated region, the excitation which
we study is concentrated in the bulk of the chain as discussed in
Sec.~\ref{sec:phase-sep} and
Figs.~\ref{fig:szi.vs.i}--\ref{fig:sz1R.vs.V.2}.  Thus we expect that
its excitation energy is not affected significantly by the boundary
conditions for sufficiently long chains.  However, in the
phase-separated region, the excitation lives near the ends of the chain
and its energy may well be strongly affected by the boundary
conditions.  In this case, the energy which we measure is still a lower
bound on the bulk gap.  This follows because the state which we study
is the lowest energy one with these quantum numbers.  If the bulk gap
were lower, we would expect a lower energy state to exist, localized
far from the chain ends.  Thus our results give strong evidence for a
spin gap for all $V>0$ but only give a reliable estimate of the size of
the gap for $V_{\rm c2}<V<V_{\rm c1}\approx 3.5t$, except for
magnitudes less than the finite-size gap as discussed above.

\subsection{Phase Transition at $V=V_{\rm c}\approx -1.5t$}
\label{sec:negV}

As discussed in Ref.~\cite{Affleck}, for small $V<0$ the pair-hopping
model is identical to the positive $U$ Hubbard model.  Thus we expect a
charge gap but no spin gap in this region.  It was also argued that
there should be a phase transition at finite $V<0$ because at
$V/t\to -\infty$ there is no charge gap but a spin gap.  In
Fig.~\ref{fig:gaps-all-V} we present DMRG results confirming this
prediction, with the transition occurring at $V=V_{\rm c}\approx
-1.5t$.  Our numerical results are consistent with the spin gap
appearing at the same critical coupling at which the charge gap
disappears; however, the presence of two distinct critical couplings
cannot be ruled out.  It is unclear to us whether this critical point
(or points) simply corresponds to the renormalized couplings $g_{\rm
s}$ and $g_3$ passing through zero, or to some more exotic critical
point.

\subsection{Phase Separation at $V=V_{\rm c1}\approx 3.5t$}
\label{sec:phase-sep}

\begin{figure}
\epsfxsize=\figwidth\centerline{\epsffile{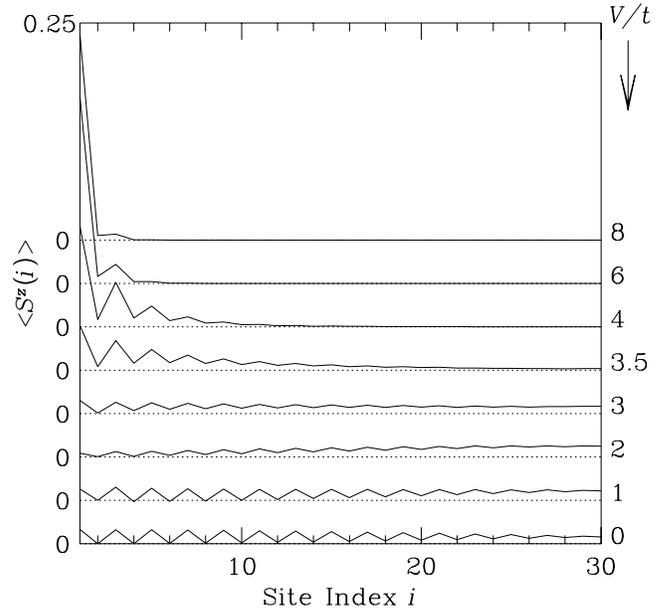}} % epsfig
\caption{Expectation values $\langle S^z(i)\rangle$ for
for different values of $V/t$,
for one electron added relative to half-filling.
The unpaired electron delocalizes
into the chain near $V/t=3.5$.
(The $L=60$ chain is symmetric about its central link.)}
\label{fig:szi.vs.i}
\end{figure}

\begin{figure}
\epsfxsize=\figwidth\centerline{\epsffile{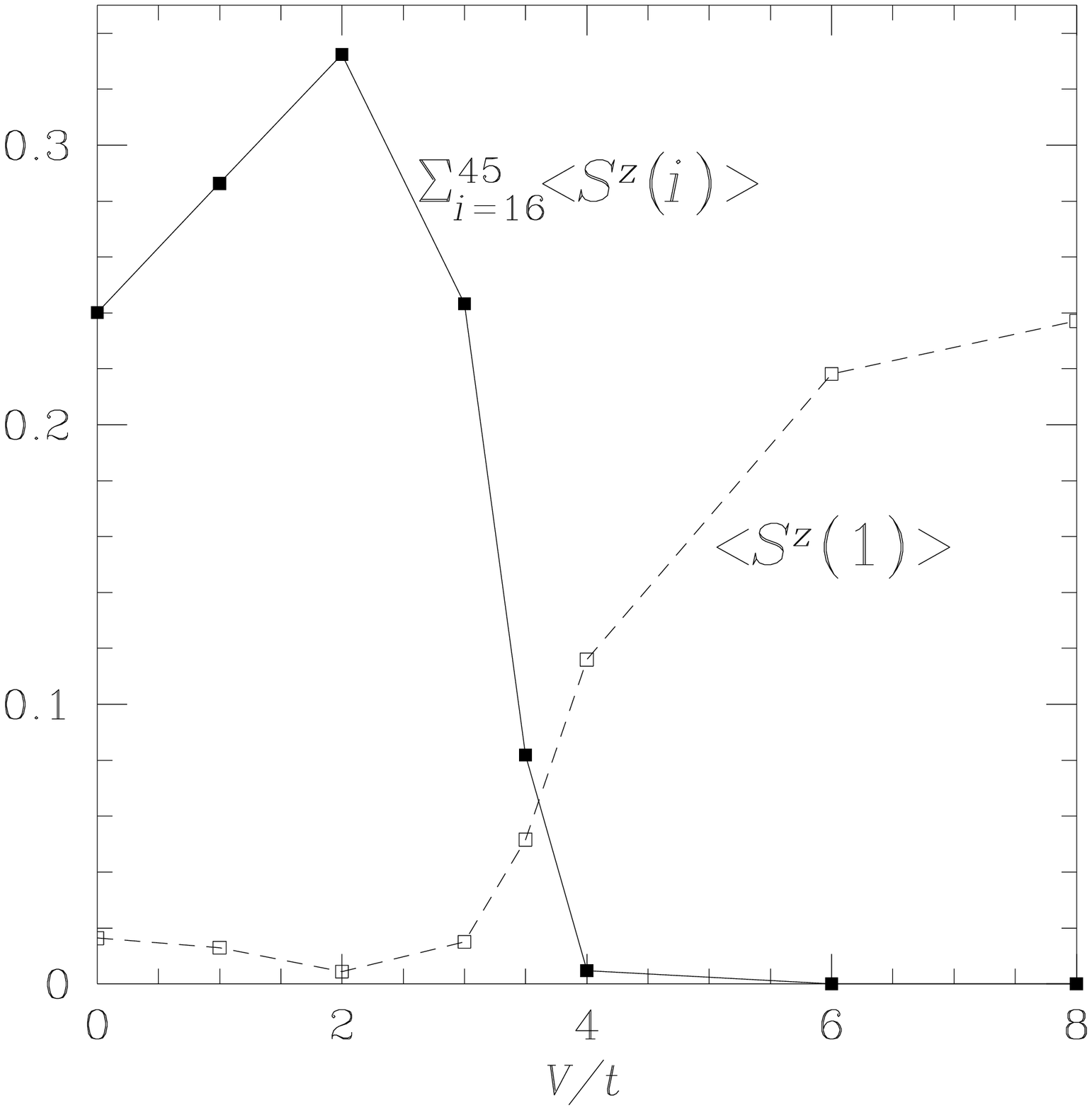}} % epsfig
\caption{Spin at a chain end (open squares) and net spin in the centre
half of the chain (filled squares) as a function of $V/t$ for a single
added electron.}
\label{fig:sz1R.vs.V.1}
\end{figure}

\begin{figure}
\epsfxsize=\figwidth\centerline{\epsffile{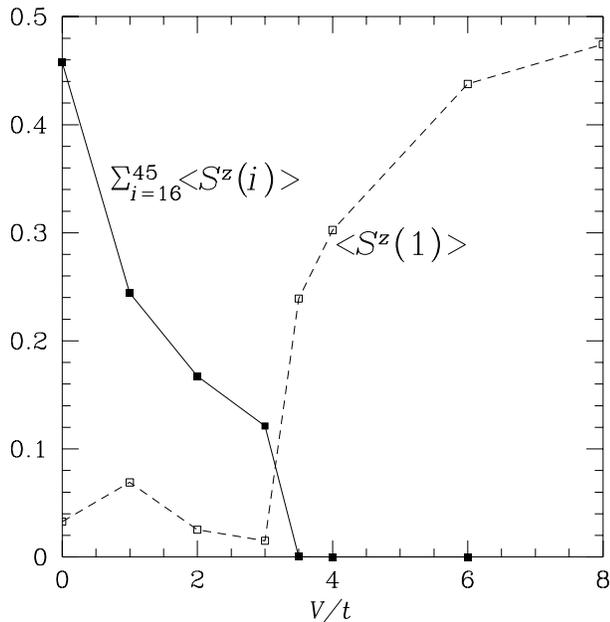}} % epsfig
\caption{Spin at a chain end (open squares) and net spin in the centre
half of the chain (filled squares) as a function of $V/t$ for two added
electrons.}
\label{fig:sz1R.vs.V.2}
\end{figure}

To demonstrate the phase-separation transition, we examine the
behaviour of wave functions obtained using the DMRG at $L=60$
in the sector of one electron added relative to half-filling.
Specifically, we plot in Fig.~\ref{fig:szi.vs.i} the
expectation value of $S^z(i)$ for sites $i=1,\ldots,30$ (the chain is
symmetric about the central link) for different values of $V/t$.  For
large $V$, the excess spin is localized at the chain ends, and as $V$
is reduced, the spin extends further into the bulk.  As $V/t$ drops
 from 4 to 3, looking at the wave function near the centre of the chain
shows that near these values of $V/t$ the spin becomes unbound from the
chain end and is rapidly and fully delocalized into the bulk of the
chain, leading us to consider $V_{\rm c1}\approx 3.5t$ as a
phase-separation critical point.  This conclusion is further verified by
examining the spin on the chain end as a function of $V/t$, as well as
the total spin in the centre half of the chain, as shown in
Figs.~\ref{fig:sz1R.vs.V.1} and \ref{fig:sz1R.vs.V.2}.

Due to the fact that we have employed the infinite system method,
instead of the finite system method\cite{White}, these wavefunctions
are not expected to be precise particularly near the phase-separation
transition and at the chain ends.  However, we expect that the results
are accurate to within a few per cent at worst, certainly not affecting
the qualitative behaviour of our figures which clearly demonstrate the
phase-separation transition.

While {\it a priori} this phase transition could occur at a different
value of $V$ than the bulk phase separation, the simplest scenario
would have both transitions occurring at the same point:  essentially
the bulk transition drives the boundary transition.  The numerical
evidence on one and two added electrons seems to indicate that for low
net magnetization, $V_{\rm c1}$ is constant.

This phase-separation transition will occur for finite $V_{\rm c1}$ in
the periodic and infinite chain as well (though not necessarily at the
same value of $V_{\rm c1}$ as for the open chain):  added unpaired
electrons will still break the chain into chainlets and the energy will
be minimized if they clump together.  However, it will be more
difficult to detect in a periodic chain since the ground state is
usually translationally invariant.

\section{Conclusions}

We conclude that there is a finite spin gap for all positive $V$ in the
half-filled pair-hopping model in one dimension, and that to accurately
describe its behaviour as a function of $V$, one must neglect the
coupling $g_4$ in the renormalization group flows.

We conclude that there are phase-separation transitions in the
pair-hopping model, one at positive $V$ and one at negative $V$.  In
one dimension at low doping from half-filling, for
$V>V_{\rm c1}\approx 3.5t$ polarized electrons clump together.

We conclude that there is a new critical point at
$V=V_{\rm c}\approx -1.5t$ at which, proceeding from weak coupling, a
spin gap opens and the charge gap closes at half-filling.

\begin{acknowledgments}

We would like to thank S.~Doniach, I.~Dzyaloshinskii, V.~Emery, A.~Hui,
D.~Huse, J.B.~Marston, H.~Schulz, and E.~S\o rensen for helpful
discussions and correspondence.  This research is supported in part by
NSERC of Canada.

\end{acknowledgments}

\end{document}